\documentclass[preprint,12pt]{elsarticle}
\usepackage{graphicx}
\usepackage{bm}
\usepackage{amssymb}
\usepackage{wasysym}
\usepackage{mathrsfs}
\usepackage{setspace}
\journal{Annals of Physics}

\begin{document}
\begin{frontmatter}

\title{Pseudogap Formation and Quantum Phase Transition in Strongly-Correlated Electron Systems}
\author{Chyh-Hong Chern\footnote{chchern@ntu.edu.tw}}
\address{Department of Physics, National Taiwan University, Taipei 10617, Taiwan}

\date{\today}
\begin{abstract}
Pseudogap formation is an ubiquitous phenomena in strongly-correlated superconductors, for example cuprates, heavy-fermion superconductors, and iron pnictides.  As the system is cooled, an energy gap opens in the excitation spectrum before entering the superconducting phase.  The origin of formation and the relevancy to the superconductivity remains unclear, which is the most challenging problem in condensed matter physics.  Here, using the cuprate as a model, we demonstrate that the formation of pseudogap is due to a massive gauge interaction between electrons, where the mass of the gauge boson, determining the interaction length scale, is the consequence of the remnant antiferromagnetic fluctuation inherited from the parent compounds.  Extracting from experimental data, we predict that there is a quantum phase transition belonging to the 2D XY universality class at the critical doping where pseudogap transition vanishes.
\end{abstract}


\end{frontmatter}

\section{Introduction}
Correlation is a quantum-mechanical patent with non-perturbative nature, which gives rise to diverse quantum phenomena.  Starting from the basic, the notion of exchange correlation states that two \emph{independent} fermions (bosons) experience an effective exchange repulsive (attractive) force when their wave functions highly overlap.  In the case of interacting particles, correlation often leads to versatile orderings, for example (anti-)ferromagneticsm, superconductorvity, and so on.  However, there remains insufficient understanding of what role the quantum correlation plays in the paramagnetic phase away from quantum criticality.  This question began to attract attentions after the behaviours of many transition-metal oxides were found outside the box of the Fermi liquid theory.  They were soon classified as \emph{strongly-correlated} materials.  The interest, with correlation still unknown, reaches its peak after the discovery of the $Cu$-based transition-metal oxides (cuprates), so-called the high-$T_c$ superconductors~\cite{Bednorz1986,wu1987}.  

The parent compounds of the high-$T_c$ superconductors is insulating antiferromagnets (AFM)~\cite{king1987}.  After chemical doping with charge carriers, the antiferromagnetic ordering disappears, followed by an enigmatic paramagnetic phase before the superconductivity emerges.  The paramagnetic phase is now known as the pseudogap phase where a gap opens in the electronic spectrum without exhibiting any signature of conventional phase transition at $T = T^*$, higher than the superconducting transition temperature $T_c$~\cite{shen2003}.  Similar phenomenon is also seen in heavy-fermion superconductors~\cite{Thompson2002} and iron pnictides~\cite{hosono2008}, which share similar phase diagram to cuprates.  This ambiguous transition is often regarded as a crossover.  After almost three decades from its discovery, the central debates still focuses on the formation of the pseudogap and its relevancy to the superconductivity in the lower temperature range.

Intensive studies have been done both experimentally and theoretically to identify whether or not the pseudogap phase belongs to a broken-symmetry state.  Recently, experimental data in cuprates are accumulated to indicate that the pseudogap phase breaks time-reversal symmetry and preserves the translational symmetry~\cite{Bourges2006,kapitulnik2008,Bourges2008,greven2008,shen2011}.  However, in most of those data, the time-reversal symmetry begins to fluctuate precursory to the pseudogap transition.  So far, the evidence is still vague that the pseudogap formation belongs to any symmetry-breaking scenario.

On the other hand, it is generally believed that the strong electron-electron repulsive interaction baptises the strong correlation.  This naive belief was first challenged by Comanac et al.~\cite{comanac2008}.  They found that a large Hubbard $U$ value is not needed but the antiferromagnetic correlation is crucial to fit the experimental data of optical conductivity.  A recent numerical calculation also indicates that Hubbard $U$ value decreases as the system size increases~\cite{Gull2013}.  Meanwhile, it has been advocated by Laughlin that the Coulomb interactions in the cuprates is simply the same as they are in elemental Si or Na metal~\cite{laughlin2013}.  However, to completely overrule the wrong belief, a mechanism responsible for pseudogap formation is needed in a framework of the \emph{weak-coupling} theory.

In this article, we construct a weak-coupling theory for the pseudogap formation, where the gauge interaction weakly coupled to electrons acquires a mass leading to a gap-like structure in the electronic spectrum.  The non-perturbative mass acquisition mechanism identifies the quantum correlation, where the remanent antiferromagnetic fluctuation becomes the longitudinal mode of the gauge field.  Moreover, the pseudogap transition is identified as a BKT-\emph{like} transition, and the transition temperature is computed.  Most importantly, we provide a scheme for the pseudogap formation \emph{without} breaking time-reversal and translational symmetry.  Finally, a generalisation to the iron pnictides and heavy-fermion systems is briefly discussed.

\section{The model}
In cuprates, mobile charge carriers are introduced in the Mott insulator by chemical doping.  As the electrons become more and more mobile, the electron scattering process becomes more and more important.  The complete description of the scattering process should include the current-current (CC) interaction in the one-band Hubbard model
\begin{eqnarray}
\mathscr{H}=-t\!\!\!\!\sum_{<ij>,\sigma}\!\!(c^\dag_{i,\sigma}c_{j,\sigma} \!&+&\! h.c.)+U_0\sum_i n_{i\uparrow}n_{i\downarrow}\nonumber \\ &+&U_1\sum_{q}\vec{J}_{\uparrow}(q)\cdot\vec{J}_{\downarrow}(-q), \label{hami}
\end{eqnarray}
where $c^\dag_{i,\sigma} (c_{i,\sigma})$ is the electron creation (annihilation) operator, $n_{i,\sigma}=c^\dag_{i,\sigma}c_{i,\sigma}$, and $\vec{J}_{\sigma}(q)$ is the current operator
\begin{eqnarray}
\vec{J}_{\sigma}(q)=\sum_p c^\dag_{q,\sigma}c_{p+q,\sigma}(\vec{p}+\frac{\vec{q}}{2}).
\end{eqnarray}
When the scattering process occurs in the lattice level, the vertex may represent a copper site.  Even though the CC interaction can be small in most strongly-correlated materials, it is a relevant ingredient for the description of the non-perturbative effect, which the pseudogap formation will be demonstrated to be one later.  Most importantly, the introduction of the CC interaction allows us to formulate a weak-coupling theory for high-$T_c$ materials, which was originally thought as a problem similar to QCD.  Taking the extended Hubbard model to the continuous limit, the theory can be written in the path integral formalism 
\begin{eqnarray}
Z &=& \int \mathscr{D}\psi^\dag\mathscr{D}\psi e^{i\int dtdx^2 \mathscr{L}}, \nonumber \\
\mathscr{L}&=& \sum_{\sigma}\psi^\dag_{\sigma}(x) (i\partial_0 +\frac{\nabla^2}{2m})\psi_{\sigma}(x) -u_0\rho_{\uparrow}(x)\rho_{\downarrow}(x)\nonumber \\&&-u_1\vec{J}_{\uparrow}(x)\cdot \vec{J}_{\downarrow}(x),\label{lagrangian1}
\end{eqnarray}
where $\hbar = 1$, $\partial_0 = \partial/\partial t$, and $\rho_\sigma(x)$ and $\vec{J}_\sigma(x)$ are the charge and the current density respectively.  The theory described by Eq.~(\ref{lagrangian1}) is, however, not renormalizable since the dimension of the coupling constants $u_0$ and $u_1$ is $[u_0]=[u_1]=L$.  This situation is similar to Fermi's $\beta$-decay theory, where a non-renormalizable \emph{four-fermion} vertex is introduced to produce a finite matrix element between a neutron, a proton, an electron, and a neutrino~\cite{fermi1934}.  The non-renormalizability hinders theorists from controlling the intrinsic infinity in the theory.  Later, the $\beta$-decay problem is completely solved by Glashow, Weinberg, and Salam~\cite{weiberg1980,salam1980,glashow1980}, who constructed the Standard Model for the electroweak interaction, where $W^\pm$ and $Z$ gauge bosons were introduced accounting for the weak interaction.

Here, we play the same trick.  We introduce a fictitious gauge field $(a_0, \vec{a})$ accounting for the effective interaction between electrons.
\begin{eqnarray}
\mathscr{L} = &\sum_{\sigma}&\psi^\dag_{\sigma}(i\partial_0)\psi_\sigma\!-\!\frac{1}{2m}[(-\frac{\vec{\nabla}}{i}-g\vec{a})\psi^\dag_\sigma][(\frac{\vec{\nabla}}{i}-g\vec{a})\psi_\sigma]\nonumber \\&-&ga_0\psi^\dag_\sigma\psi_\sigma\!-\!\frac{1}{4}f_{\mu\nu}f_{\mu\nu}\!-\!\frac{1}{2}M_1^2\vec{a}\cdot\vec{a}\!+\!\frac{1}{2}M_0^2a^2_0,\label{effective}
\end{eqnarray}
where $f_{\mu\nu}=\partial_\mu a_\nu - \partial_\nu a_\mu$ is the field strength, $g$ is the coupling constant, $M_0$ and $M_1$ are the mass parameters for the $a_0$ and $\vec{a}$ component respectively, and the metric $(1, -1, -1)$ is adopted.  Since the theory is non-relativistic, there are two mass parameters for different interaction strengths in the charge and the current channels.  Integrating over $a_0$ and $\vec{a}$, Eq.~(\ref{effective}) becomes
\begin{eqnarray}
\mathscr{L}= \sum_{\sigma}&\psi^\dag_{\sigma}&(x) (i\partial_0 +\frac{\nabla^2}{2m})\psi_{\sigma}(x) \nonumber \\&+&\frac{-ig^2}{2}\sum_{\sigma,\sigma'}\rho_{\sigma}(x)\frac{i}{k^2-M_0^2+i\eta}\rho_{\sigma'}(x)\nonumber\\&-&\frac{-ig^2}{2}\sum_{\sigma,\sigma'}\vec{J}_{\sigma}(x)\frac{i}{k^2-M_1^2+i\eta}\cdot \vec{J}_{\sigma'}(x), \label{lagrangian2}
\end{eqnarray}
where $k^2=k_0^2-\vec{k}^2$, $\eta$ is a small parameter, and terms with higher order of $O(g^4)$ are ignored.  Using the Grassmann variable, terms with the same spin at the same spatial location automatically vanish.  Comparing Eq.~(\ref{lagrangian1}) and Eq.~(\ref{lagrangian2}), $u_0$ and $u_1$ can be obtained
\begin{eqnarray}
u_0=\frac{g^2}{M_0^2}, \ \ u_1=-\frac{g^2}{M_1^2} \label{couplingconstant}
\end{eqnarray}
in the low-energy and long-wavelength limit.  Now, the coupling constant $g$ has a dimension $[L^{-1/2}]$.  The theory becomes renormalizable.  Moreover, $u_0 > 0$ represents the repulsive interaction, and $u_1 < 0$ represents the attractive interaction.  It is consistent with our physical intuition of electromagnetism.  Namely, the charge interaction between two electrons is repulsive due to the Coulomb interaction.  The current interaction is attractive since two conducting wires with the same direction of electric current attract to each other due to the Lorentz force.  Similar situation of the current interaction between spinons was considered by Lee {\emph et al.} before~\cite{lee2007}.  Different from other mechanism by exchanging vector bosons, for example phonons or magnons, the exchange of gauge boson produces both repulsive and attractive interactions.

\section{Mass acquisition of the gauge boson}
Mass of a gauge boson determines the interaction length scale since $M^{-1}$ has the dimension of length.  However, the finiteness of mass makes Eq.~(\ref{effective}) not gauge-invariant.  As we know, the parent compound of cuprates is a 3D antiferromagnet (AFM)~\cite{king1987}.  In addition, the spin anisotropy favours the directions of spins to point in the cooper-oxygen plane~\cite{king1987,oka1988}.  By the chemical doping, 3D AFM vanishes and antiferromagnetic (AF) correlation becomes short-ranged and highly anisotropic in space~\cite{Schneider2007,auerbach2009}.  Due to the spin anisotropy, the remnant AF fluctuation can be described by a phase field $\phi (t, \vec{x})= (1/q)e^{i\sigma(t, \vec{x})}$.  Now, as two electrons interact with each other by exchanging a gauge boson, in the path of the exchange, the gauge boson couples to the AF fluctuation, which can be described by the following Lagrangian
\begin{eqnarray}
\mathscr{L}_{S} &=& \frac{1}{2}M_0^2(D_0 \phi)^\dag(D_0 \phi)-\frac{1}{2}M_1^2(D_i \phi)^\dag(D_i \phi) \nonumber \\ &=& \frac{1}{2}M^2_0(\frac{1}{q}\partial_0\sigma + a_0)^2-\frac{1}{2}M^2_1(\frac{1}{q}\vec{\nabla}\sigma-\vec{a})^2,
\end{eqnarray}
where $D_0 = i\partial_0 - qa_0$ and $D_i = -i\partial_i - qa_i$ are the covariant derivative, and $q$ is the gauge coupling for the phase field.  The gauge transformation is defined by
\begin{eqnarray}
\vec{a}\rightarrow\vec{a}'& =& \vec{a} + \frac{1}{q}\vec{\nabla}\lambda \nonumber \\
a_0\rightarrow a'_0 &=& a_0 -\frac{1}{q}\partial_0\lambda \nonumber \\
\sigma\rightarrow \sigma' &=& \sigma + \lambda. \label{gaugetransformation}
\end{eqnarray}
Then, the total Lagrangian is manifestly gauge invariant
\begin{eqnarray}
\mathscr{L} = \sum_{\sigma}&\psi^\dag_{\sigma}&(i\partial_0)\psi_\sigma-\frac{1}{2m}[(-\frac{\vec{\nabla}}{i}-g\vec{a})\psi^\dag_\sigma][(\frac{\vec{\nabla}}{i}-g\vec{a})\psi_\sigma]]\nonumber \\&-&ga_0\psi^\dag_\sigma\psi_\sigma-\frac{1}{4}f_{\mu\nu}f_{\mu\nu}\nonumber \\ &+&\frac{1}{2}M_0^2(D_0 \phi)^\dag(D_0 \phi)-\frac{1}{2}M_1^2(D_i \phi)^\dag(D_i \phi).\label{fulllagrangian}\end{eqnarray}
The mass acquisition of the gauge boson described here is known as the St\"uckelberg mechanism~\cite{s1938a,s1938b,s1938c}, which is a common method in the mass generation in the context of the string theory~\cite{ho2012}. 

\section{Pseudogap transition}
The Lagrangian in Eq.~(\ref{fulllagrangian}) contains the correct description for the pseudogap transition.  In the high temperature region, the gauge field is massless.  Since the phase field is fluctuating, the vacuum expectation value $<\phi> = 0$.  The phase field is like a needle in a clock.  The needle has a finite length, but different directions cancel one another in the high temperature region.  As the temperature goes down, the field tends to be static.  The vacuum expectation value of the field is frozen to $<\phi> = (1/q)e^{i\sigma_0}$~\cite{Uimin1985}.  Now, we can simply gauge away the phase field by choosing $\lambda = -\sigma_0$.  In this case, $<\sigma'> = 0$ and $<\phi'> = 1/q$.  The last two terms in Eq.~(\ref{fulllagrangian}) soon become the last two terms in Eq.~(\ref{effective}), and the gauge bosons acquire a mass.  Similar to the Anderson-Higgs mechanism, the gauge boson combines with the Goldstone mode of the phase field and becomes massive.  In fact, the St\"uckelberg mechanism was sometimes regarded as a special Higgs mechanism with an infinite Higgs mass. 

The phase transition described above belongs to the 2D XY universality class~\cite{Uimin1985,fisher1990}.  Therefore, a Berezinski-Kosterlitz-Thouless (BKT) transition at finite temperature is expected~\cite{thouless1973,fisher1990}.  The transition temperature is $T^* = (\pi M_1^2)/(2q^2)$ for the simplest case when $M_0\approx M_1$.  The BKT transition is a phase transition of infinite order.  In the high temperature phase, it is a disordered phase with an exponential correlation, and in the low temperature phase, it is a quasi-ordered phase with a power-law correlation.  In this case, the pseudogap is a BKT-\emph{like} transition.  In the low temperature phase, the AF fluctuation becomes the longitudinal mode of the gauge boson.  In the unitary gauge, the propagator of the $\sigma(t, \vec{x})$ field is simply zero.  Namely, it does not appear in any physical process, nor can the power-law AF correlation be observed.  

Let us now consider the symmetry property in the pseudogap state.  Since $a_0$ is even and $\vec{a}$ is odd under time-reversal transformation, the phase field $\phi \rightarrow \phi*$ under time-reversal transformation.  In the high temperature phase, $<\phi>=0$, so it is time-reversal symmetric.  In the low temperature phase, the phase field can always be gauged away and becomes $<\phi>=<\phi*>=1/q$.  Both time-reversal symmetry and translational symmetry is preserved in the pseudogap phase.

\begin{figure}[htbp]
\begin{center}
\includegraphics[scale=0.24]{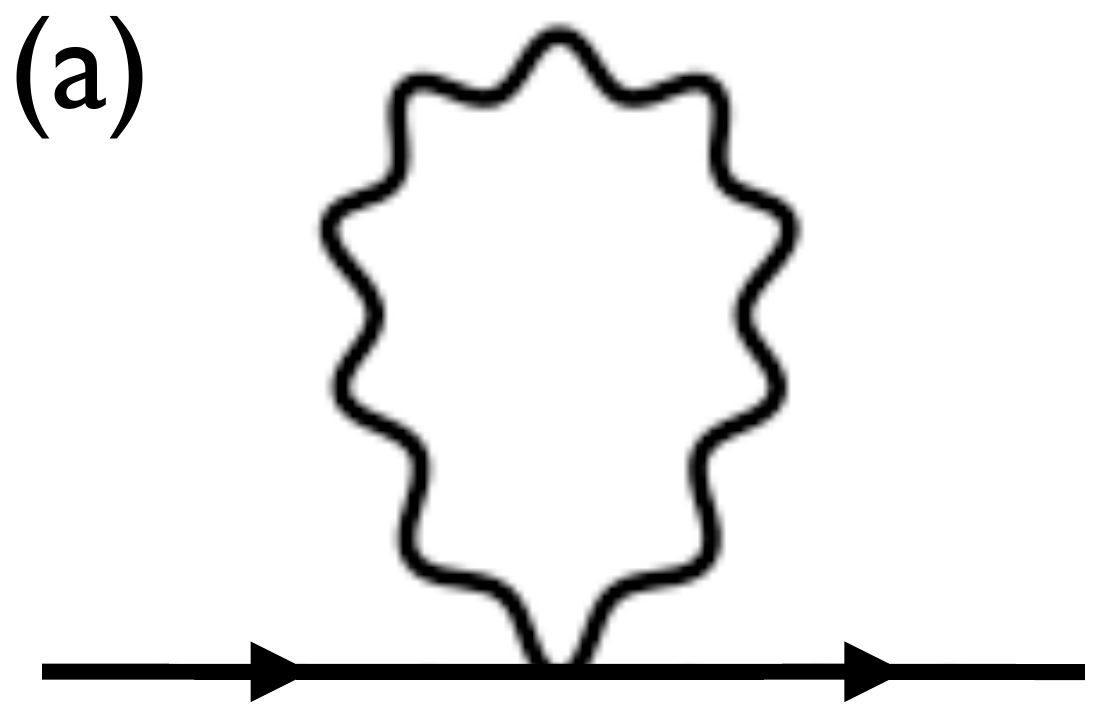}
\includegraphics[scale=0.24]{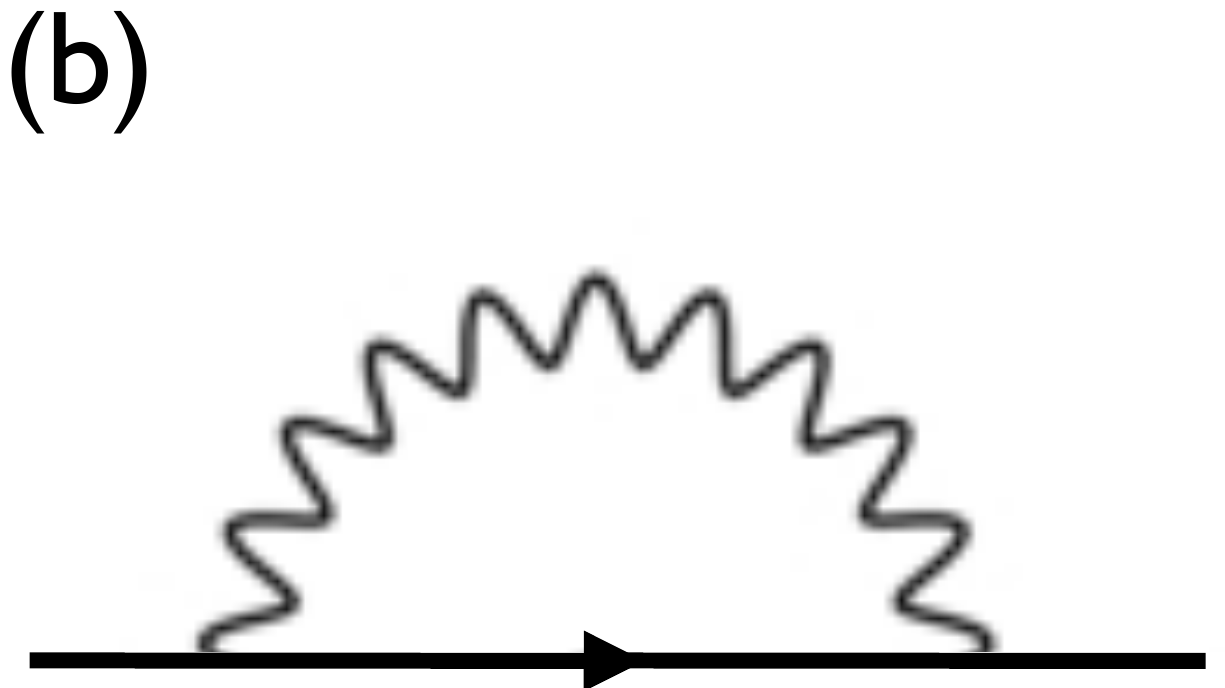}
\caption{Feynman diagrams to compute the self energy of electron}
\label{fig2}
\end{center}
\end{figure}

Across the pseudogap transition, the band structure is also modified.  Using the standard Feynman diagram technique, we consider the diagrams up to $O(g^2)$ as shown in (Fig. 2) for the computation of the electron self-energy.  We note that the dimensionless vertex $g^2/2m$ in (Fig. 2A) has the value $10^{-5}$ in support of the current perturbative scheme.  Given the full Green's function of electron $G^{-1}(\omega,\vec{p}) = \omega - \varepsilon_p - \Sigma(\omega, \vec{p}) + i\eta$, 
\begin{eqnarray}
\Sigma(\omega, \vec{p}) = \Sigma_1(\omega, \vec{p})+\Sigma_2(\omega, \vec{p}),
\end{eqnarray}
where $\Sigma_1(\omega, \vec{p})$ and $\Sigma_2(\omega, \vec{p})$ are the amplitudes for the diagrams in (Fig. 2a) and (Fig. 2b) respectively.  After some algebra, $\Sigma_1(\omega, \vec{p})$ = $(g^2/2m)(M_1/4\pi+M^2_0/(12\pi M_1))$ and Re$\Sigma_2(0, 0) = 0$.  Namely, the diagram in (Fig. 2a) contributes to a finite energy gap and the one in (Fig. 2b) does not.  We can do the same calculation for the hole in the valence band.  Then, we obtain a finite energy gap 
\begin{eqnarray}
\Delta = \frac{g^2}{4\pi m}(M_1+\frac{1}{3}\frac{M^2_0}{M_1}),
\end{eqnarray}
which vanishes when $M_0=M_1=0$.  Therefore, at the pseudogap transition, the gauge field combines with the AF fluctuation and acquires a mass, opening a finite energy gap in the electronic spectrum, which is the origin of the pseudogap formation. 

\section{Experimental relevancy}
The current theory is a zero-temperature calculation, and the pseudogap is isotropic in momentum space.  In the real materials, the pseudogap structure can be versatile due to the competition with the superconducting state with exotic pairing symmetry.  Recently, with considerably-improved sample quality, a \emph{nodeless} pseudogap structure was confirmed experimentally in the highly-underdoped La$_{2-x}$Sr$_{x}$CuO$_{4}$ (LSCO) systems~\cite{Razzoli2013}.  While the pseudogap is nodeless in the low and zero temperature, the \emph{Fermi-arc} feature is restored and robust as either temperature or doping level increases~\cite{Razzoli2013,shen2007,Damascelli2010}.  The interesting temperature and doping dependence of the pseudogap structure might be due to the momentum-dependent hierarchy energy scales~\cite{rice2000}.  The pseudogap in cuprate is seemingly intrinsically nodeless, since even in La$_{2-x}$Ba$_{x}$CuO$_{4}$ ($x=1/8$), the failed high-$T_c$ superconductor, a gap is opened in the nodal direction when the system is approaching to zero temperature~\cite{shen2009}.  While the robustness of the Fermi arc at finite temperature remains intriguing, further studies with the consideration of the lattice symmetry and the $d$-wave superconducting instability should clarify this issue.

Estimating the parameters in the theory, the fundamental properties in cuprates can be reproduced.  Taking $M_0 \approx M_1 \approx 2.5\times 10^3$ eV, $g^2 \approx 50$ eV, and $q^2 \approx 4\times 10^8$ eV, we obtain the Hubbard term $U\approx 8$ eV, the pseudogap magnitude $\Delta \approx 20$ $m$eV~\cite{Razzoli2013,shen2007,shen2009,Damascelli2010}, and the pseudogap transition temperature $T^* \approx 1.5\times10^2$~K.  The values of $M_0$ and $M_1$ imply the length scale $\sim \AA$, which explains why the on-site interaction is most relevant.  As observed in the angle-resolved photoemission, the magnitude of the pseudogap and the pseudogap transition temperature has a roughly linear relation with doping~\cite{shen2003}.  Assuming that the Hubbard-$U$ value is correlated with the magnitude of the pseudogap, we can obtain the doping dependence of the $M$, $g^2$ and $q^2$ phenomenologically as shown in (Fig. 3).  While the interaction length scale is independent of doping, the electron-electron interaction decreases monotonically as the doping increases.  At the critical doping $x=x_c$, $g^2$ is zero and the system becomes Fermi liquid.  Interestingly, we found that the amplitude of the AF fluctuation, $|\phi|$, also vanishes at $x_c$ with $|\phi| \sim (x_c-x)^{1/2}$ provided that $T^* \sim (x_c-x)$.   A quantum phase transition is implied as the onset of the AF fluctuation.  Since it is the end point of the finite-temperature pseudogap transition, the quantum phase transition belongs to the 2D XY universality class.  In other words, the quantum phase transition is a continuous phase transition of infinite order.

\begin{figure}[htbp]
\begin{center}
\includegraphics[scale=0.18]{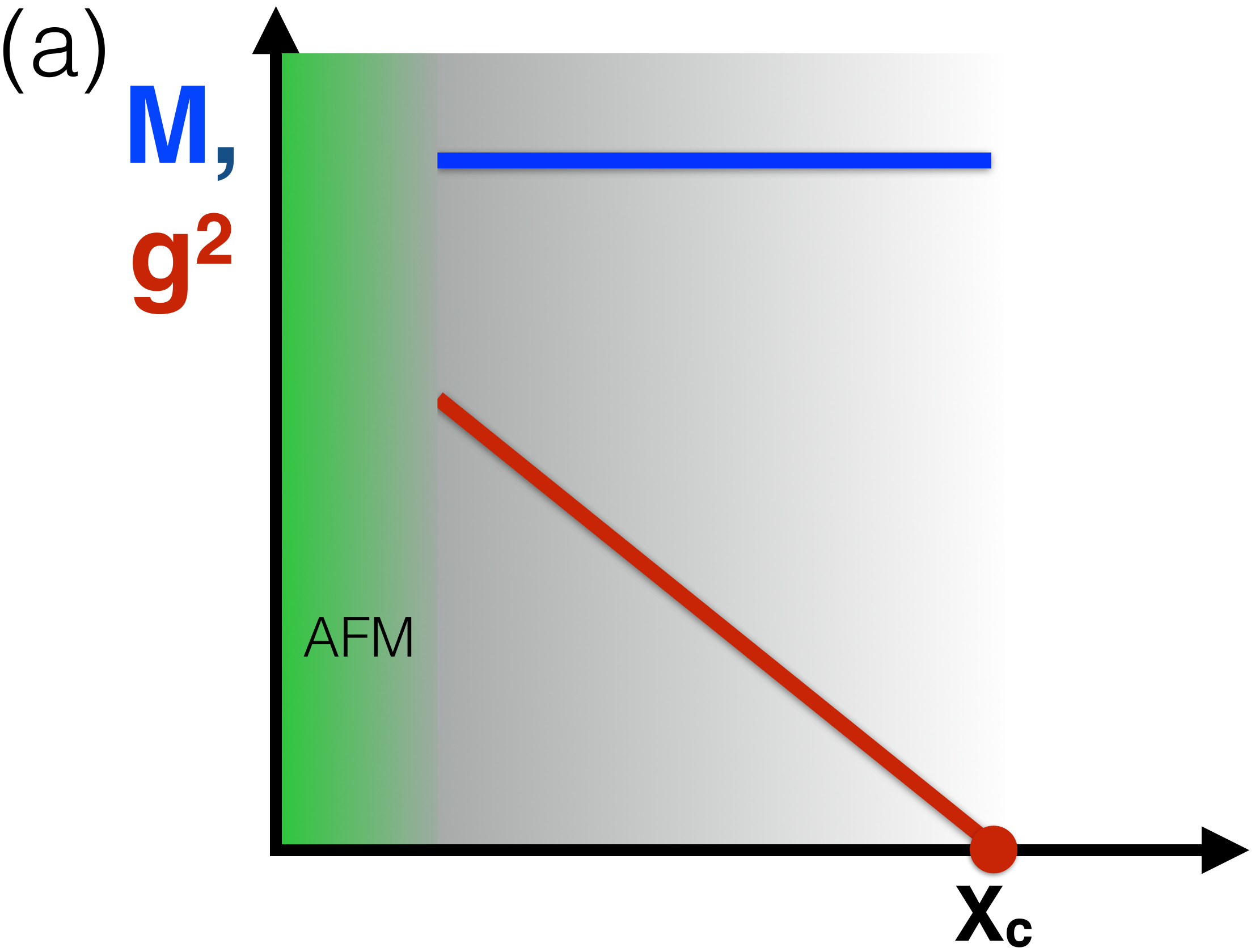}
\includegraphics[scale=0.18]{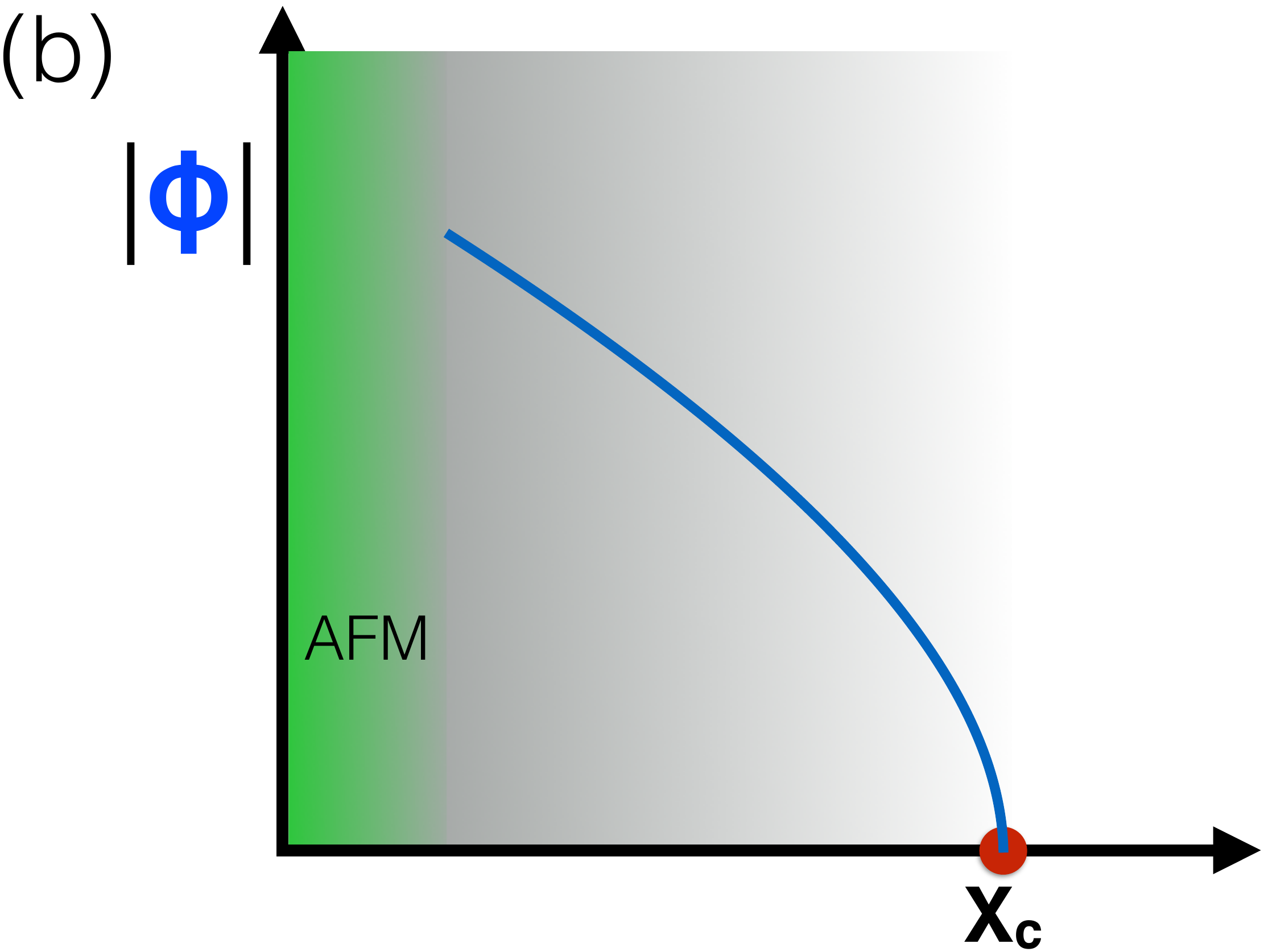}
\caption{Schematic doping dependence of the $M$ ($M_0 \approx M_1$ is assumed. labelled in blue in (a)), the gauge coupling $g^2$ (labelled in red in (a)), and the amplitude of the AF fluctuation $|\phi|$ (labelled in blue in (b)). $x_c$ represents the critical doping where the quantum phase transition occurs. The gradient green area represents the antiferromagnetic phase.  The gradient grey area is the pseudogap phase.  }
\label{fig3}
\end{center}
\end{figure}

\section{Conclusion and summary}
Introducing the massive gauge interaction, our approach offers a detour from the Hubbard model toward a weak-coupling theory in the continuous limit.  The current theory can also describe the interaction of finite range, using different mass scales, which can be applicable to iron pnictides and heavy-fermion superconductors~\cite{lee2013}.  Nevertheless, it sheds light on the pseudogap formation in those systems.  Similar to cuprates, the existence of the remnant AF phase fluctuation could be universal, after the AFM is frustrated by chemical doping or pressure.  Different from cuprates, iron pnictides and heavy-fermion superconductors are multi-band systems.  Considering the extended Hubbard model of the minimal active bands, a natural generalization is to introduce a non-abelian gauge interaction.  Similar to the $W^\pm$ and $Z^0$ bosons in the weak interaction, non-abelian gauge interactions in the multi-band systems, coupled to the $U(1)$ AF fluctuation, allow different mass (or zero mass) depending on the magnitudes of the pseudogap (or no gap).  For further development, more experimental constraints are required, for example, whether or not there are multiple pseudogap transitions at different temperatures and there are different magnitudes of the pseduogap?  Finally, we remark that the current theory does not imply that the real photon acquires a mass in the pseudogap phase.  The real photon couples to the electron but does not couple to the AF fluctuation since the phase field already becomes the longitudinal mode of the internal gauge boson.

Different from other gauge theories in the strong-coupling limit~\cite{patrick2006}, we construct a weak-coupling gauge theory for the pseudogap formation in cuprates without breaking time-reversal and translational symmetry.  It is not electrons but the \emph{interaction} between electrons that couples to the AF fluctuation.  The non-perturbative mechanism of the mass acquisition identifies the strong correlation in cuprates.  The pseudogap formation is, therefore, a signature of the strong correlation.  The strong electron-electron interaction and the strong correlation should \emph{not} be on the equal footing.  The relevancy of the pseudogap phase and quantum phase transition to the exotic superconductivity should be the next radical research direction to pursue. 

\section{Acknowledgement}
I deeply appreciate the enlightening discussions with Pei-Ming Ho, who patiently taught me the St\"uckelberg physics.  I am grateful for the stimulating discussions with Jiunn-Wei Chen.  For many years, with highest gratitude, my high-$T_c$ knowledge had been benefited from the educational discussions with Robert Laughlin, Dung-Hai Lee, Naoto Nagaosa, and Shoucheng Zhang.  This work was supported by the National Science Council of Taiwan under NSC 100-2112-M-002-015-MY3, NTU Grant No. 101R7831, and Center for Theoretical Sciences in NTU. 

%
\end{document}